\documentclass[preprint,showpacs,preprintnumbers,amsmath,amssymb]{revtex4}

\usepackage{graphicx}
\usepackage{dcolumn}
\usepackage{bm}
\usepackage{url}%

\begin{document}

\title{A New calculation of the atmospheric neutrino flux in a 3-dimensional scheme.}

\author{M.~Honda}
\email[]{mhonda@icrr.u-tokyo.ac.jp}
\homepage[]{http://icrr.u-tokyo.ac.jp/~mhonda}
\author{T.~Kajita}
\email[]{kajita@icrr.u-tokyo.ac.jp}
\affiliation{Institute for Cosmic Ray Research, University of Tokyo, 
Kashiwa-no-Ha, Chiba 277-8582, Japan}
\author{K.~Kasahara}
\email{kasahara@sic.shibaura-it.ac.jp}
\affiliation{Shibaura Institute of Technology, Fukasaku, Ohmiya, Saitama 330-8570, Japan}
\author{S.~Midorikawa}
\email{midori@aomori-u.ac.jp}
\affiliation{Faculty of Engineering, Aomori University, Aomori 030-0943, Japan.}

\date{\today}

\begin{abstract}
We have revised the calculation of the flux of atmospheric neutrinos
based on a 3-dimensional scheme with the realistic 
IGRF geomagnetic model. 
The primary flux model has been revised, based on the AMS 
and BESS observations, and the interaction model updated to DPMJET-III.
With a fast simulation code and computer system, 
the statistical errors in the Monte Carlo study are negligible.
We estimate the total uncertainty of the atmospheric neutrino 
flux prediction is reduced to $\lesssim$~10~\% below 10~GeV.
The `3-dimensional effects' are found to be almost the same as the
study with the dipole magnetic field,
but the muon curvature effect remains up to a few tens of GeV for 
horizontal directions.
The uncertainty of the absolute normalization of the atmospheric 
neutrino is still large above 10~GeV due to the 
uncertainty of the primary cosmic ray flux above 100~GeV.
However, the zenith angle variation is not affected by these 
uncertainties.
\end{abstract}

\maketitle
\section{\label{sec:intro}Introduction}

The discovery of the neutrino oscillation from the study of atmospheric 
neutrinos is a one of the most important results in recent
physical research~\cite{sk} 
(see also Refs.~\cite{fukuda,imb,soudan2,macro}, 
and Ref.~\cite{KT} for a review).
The study is carried out by the comparison of theoretical calculation
of the atmospheric neutrino flux and experimental data.
Therefore, it is desirable that both theoretical and 
experimental studies are improved.
The SuperKamiokande is improving the statistics
and the accuracy steadily for experimental data.
It is important to improve the theoretical prediction of the
atmospheric neutrino flux also.

There have been some improvements in the theoretical prediction of
the atmospheric neutrino flux\cite{hkkm95,gaisser-new,fluka-battis,
engel-hamburg, hkkm-hamburg, hkkm-tsukuba}
(see Ref.~\cite{gaisser-honda} for a review).
These studies were useful to determine the flux ratios between 
different types of neutrinos and the variation over zenith angle 
with good accuracy, 
and establish neutrino oscillations and the existence of neutrino masses.
We now wish to improve the accuracy of the absolute normalization
as well as the ratio and directionality of the atmospheric neutrino 
fluxes for further studies.

In the time since our last comprehensive study of the atmospheric neutrino flux~\cite{hkkm95},
knowledge of the primary cosmic ray has been improved by 
observations such as BESS~\cite{bess} and AMS~\cite{ams} below 100~GeV.
There have also been theoretical developments in hadronic 
interaction models such as Fritiof 7.02~\cite{fritiof7.02}, 
FLUKA97~\cite{fluka} and DPMJET-III~\cite{dpmjet3}.
Here, we adopt these revised primary flux and hadronic interaction models.

It has also been pointed out that the atmospheric neutrino flux 
calculated in a 3-dimensional scheme is significantly different 
from that calculated in a 1-dimensional scheme at low energies for  
near horizontal 
directions~\cite{fluka-battis,lipari-ge,lipari-ew,hkkm-dipole,wentz,liu,bartol-3d}.
The 1-dimensional approximation has been widely used
in the past, and was used in our previous calculation and others~\cite{hkkm95,gaisser-new}.
This approximation is justified by the nature of hadronic interactions for  
calculations of high energy ($\gtrsim$~10~GeV) atmospheric neutrino fluxes, 
but not at lower energies.
With the computer resources then available, however, it was difficult to 
complete the calculation of atmospheric neutrino fluxes in a full 3-dimensional 
scheme within a tolerable length of the time.
Some of 3-dimensional calculations 
employ approximations based on symmetry to circumvent 
the impact of limited computer resources.
In Ref.~\cite{fluka-battis}, spherical symmetry is assumed, ignoring the
magnetic field in the atmosphere, and in
our previous 3-dimensional calculation~\cite{hkkm-dipole} we assumed an
axial symmetry and used a dipole geomagnetic field model.
Thus, 
a detailed calculation in a full 3-dimensional scheme without
symmetry remains a challenging job.

We have developed a new and fast simulation code for the propagation of 
cosmic rays in the atmosphere to 
calculate the atmospheric neutrino flux in a full 3-dimensional scheme without 
having to assume symmetry.
This fast simulation code and a fast computation system allow us
to calculate the atmospheric neutrino flux with good accuracy
over a wide energy region from 0.1 to a few tens of GeV, as is shown in this paper. 
The differences between 3-dimensional and 1-dimensional calculation schemes
are similar to that we found in the study with a dipole geomagnetic 
field~\cite{hkkm-dipole}, and are small above a few GeV.
The neutrino flux calculated in the 3-dimensional scheme is 
smoothly connected to the one calculated in the 1-dimensional scheme
at a few tens of GeV.
We are therefore able to 
discuss the atmospheric neutrino flux up to 10~TeV in this paper.

Although progress in our theoretical study of the atmospheric neutrinos flux
has been reported partly elsewhere~\cite{hkkm-hamburg,hkkm-tsukuba}, 
this is the first comprehensive report since 1995~\cite{hkkm95}.

\section{\label{sec:primary} Primary cosmic ray flux model}

\begin{figure}[tbh]
\includegraphics[width=10cm]{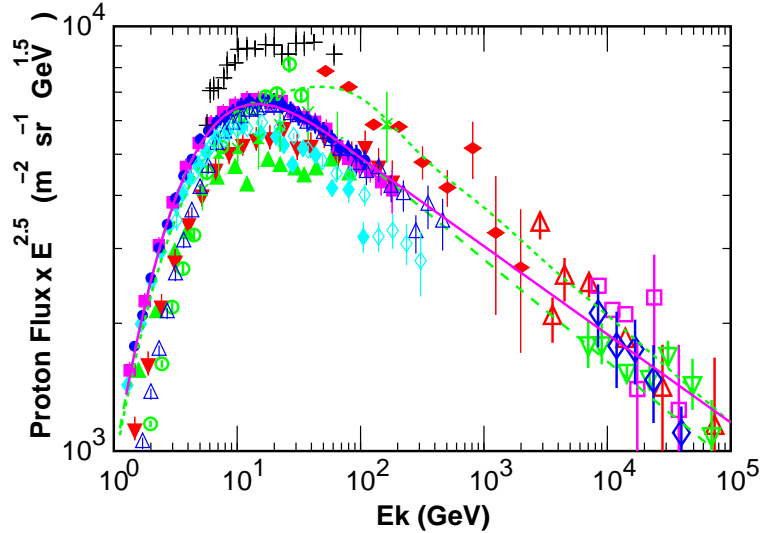}%
\caption{\label{fig:primary-proton}
Primary cosmic ray observation and our model curves for protons
at solar minimum. 
Crosses indicate data from Ref.~\cite{webber},
open circles indicate data from MASS~\cite{mass},
closed upward triangles LEAP~\cite{leap},
closed downward triangles IMAX~\cite{imax},
closed vertical diamonds CAPRICE-94~\cite{caprice94}, 
open vertical diamonds CAPRICE-98~\cite{caprice98}, 
closed circles BESS~\cite{bess},
closed horizontal diamonds AMS~\cite{ams},
open upward triangle BESS-TeV~\cite{besstev},
closed horizontal diamonds Ryan et al~\cite{ryan},
open downward large triangles JACEE~\cite{jacee-p},
open diamonds large RUNJOB~\cite{runjob},
open upward large triangles from Ivanenko et al.~\cite{moscow},
and open large squares from Kawamura et al.~\cite{kawamura}.
The dashed line shows the spectrum calculated with Eq.~\ref{eq:Todor} 
and Table~\ref{tab:Todor}, and the solid line shows that with 
the modification explained in the text.
The dotted line is the proton flux model we used in 
Ref.~\cite{hkkm95}.
}
\end{figure}

The primary flux model we use is based on the one presented 
in Refs.~\cite{gaisser-hamburg} and~\cite{gaisser-honda},
in which the primary cosmic ray data below $\sim$~100~GeV 
are compiled and parameterized with the fitting formula:

\begin{equation}
\phi(E_k)\;=\;K\times
\left(E_k\,+b\,\exp\left[-c\sqrt{E_k}\right]\right)^{-\alpha}\ .
\label{eq:Todor}
\end{equation}
where $\alpha, k, b, c$ are the fitting parameters.
Although using the same fitting formula,
the fitting parameters for nuclei heavier than helium are different 
in Refs.~\cite{gaisser-hamburg} and~\cite{gaisser-honda}.
The parameters we used are taken from 
Ref.~\cite{gaisser-honda} and tabulated in table~\ref{tab:Todor}.

\begin{table}[!htb]
\caption{Parameters for all five components 
in the fit of Eq. \protect\ref{eq:Todor}.}
\vspace*{2truemm}
\begin{tabular}{@{}|c| c c c c|@{}}
\toprule
parameter/component&$\alpha$&$K$&\hspace{5mm}$b$\hspace{5mm}&\hspace{5mm}$c$\hspace{5mm}\\ 
\colrule 
 Hydrogen (A=1) &\ 2.74$\pm$0.01\ &\ 14900$\pm$600\ & 2.15  & 0.21 \\
 He (A=4)       &  2.64$\pm$0.01  &    600$\pm$30   & 1.25\ & 0.14 \\ 
 CNO (A=14)     &  2.60$\pm$0.07  &   33.2$\pm$5    & 0.97  & 0.01 \\
 Mg--Si (A=25)  &  2.79$\pm$0.08  &   34.2$\pm$6    & 2.14  & 0.01 \\
 Iron (A=56)    &  2.68$\pm$0.01  &   4.45$\pm$0.50 & 3.07  & 0.41 \\ 
\botrule
  \end{tabular}
\label{tab:Todor} 
\end{table} 

However, the extension of this flux model for cosmic ray protons does 
not agree with emulsion chamber experiments above $\sim$~10~TeV
(the dashed line in Fig.~\ref{fig:primary-proton}).
Therefore, we modified the power index above 100~GeV to $-$2.71,
so that the fit passes through the center of the emulsion 
chamber experiments data (the solid line in Fig.~\ref{fig:primary-proton}).
We also show the flux model for cosmic ray protons used in Ref.~\cite{hkkm95} 
as the dotted line in Fig.~\ref{fig:primary-proton}.
Other than the cosmic ray protons, we use the same flux model as
Ref.~\cite{gaisser-honda}. 

Note, we employ the superposition model for the  cosmic ray nuclei, 
i.e., we consider a nucleus as the sum of individual nucleons, 
$Z$ protons and $A-Z$ neutrons.
The validity of the superposition model was discussed in Ref.~\cite{jengel}
based on the Glauber formalism of nucleus--nucleus collisions~\cite{glauber}.
The authors showed the interaction mean-free-path of a nucleon in a nucleus 
is the same as a free nucleon, and concluded that the superposition 
model is valid for the calculation of time averaged quantities, 
such as the fluxes of atmospheric neutrinos and muons.
A similar discussion was also presented in Ref.~\cite{hkkm95} with 
the same conclusion.

\section{\label{sec:intmodel}Hadronic interaction}

For the hadronic interaction model,
we are using theoretically constructed models 
which have been successfully applied to detector simulations in high energy accelerator 
experiments.
In Ref.~\cite{hkkm95}, 
we used NUCRIN~\cite{nuc1}  for 0.2~GeV $\le E_{lab} \le$ 5~GeV, 
FRITIOF version 1.6~\cite{fritiof1.6} for $ 5~{\rm GeV} \le E_{lab} \le 500~{\rm GeV}$,
and an original code developed by one of us~\cite{kasahara} 
was used above 500~GeV.
There were almost no improvements in the experimental study of the 
hadron interaction model of the multiple production, 
but there are noticeable improvements in the theoretical study, resulting in 
Fritiof 7.02~\cite{fritiof7.02}, FLUKA97~\cite{fluka}, and 
DPMJET-III~\cite{dpmjet3}. 

\begin{figure}[tbh]
\includegraphics[width=10cm]{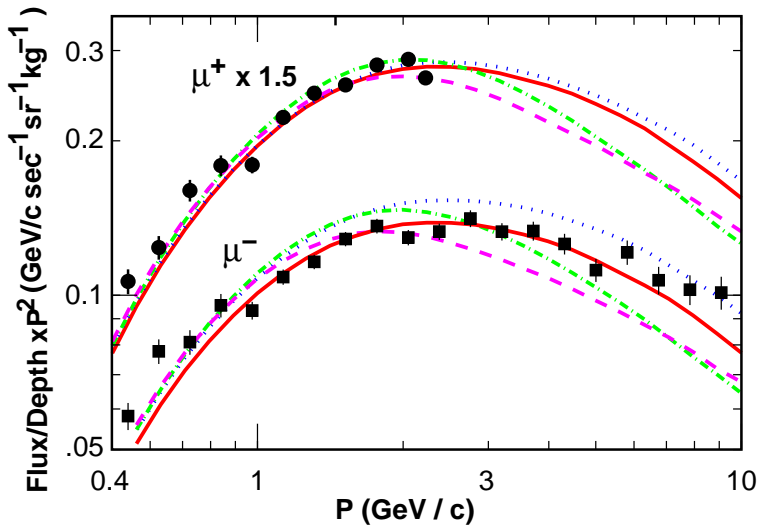}%
\caption{\label{fig:abe}
The quantity $[Flux/Depth]$ averaged over all the muon observation by BESS 2001~\cite{abe2}
at balloon altitudes.
The lines are the same quantities calculated by DPMJET-III (Solid line), 
Fritiof 1.6 (Dashed line), Fritiof 7.02 (dotted line), and FLUKA 97 (Dash dot), 
}
\end{figure}

To determine which is the better interaction model, 
we have used data on secondary cosmic ray
muons~\cite{mass-himu,mass-himu2,caprice-himu,caprice-himu2,
bess-himu,yamamoto,abe1}
and gamma-rays~\cite{bets} at balloon altitudes.
The secondary cosmic rays at the balloon altitude are ideal 
for the study of the interaction model.
They are approximately proportional to the air depth, 
and the ratio $[Flux/Depth]$ is determined almost only by 
the interaction and the flux of primary cosmic rays.
On the other hand, the small statistics due to the small flux of secondary 
cosmic rays at 
balloon altitudes is the disadvantage for this study.
The BESS 2001 flight is unique in this regard, as it measured the primary
cosmic ray and muon fluxes simultaneously a little deeper 
in the atmosphere (4--30~g$/$cm$^2$) than normal long duration flights,
and collected a sufficient number of muons and primary protons.
In Fig.\ref{fig:abe}, we show the study made for muons observed by
BESS 2001~\cite{abe1}.
Although it is hard to discriminate between other interaction models, 
it is found the DPMJET-III gives the best agreement between calculation 
and observation (for details, see Ref.~\cite{abe2}).
Note that the momentum range shown in Fig.~\ref{fig:abe} is
from 0.4 to 10~GeV/c.
and that primary cosmic rays with energies from 6 to 80~GeV are 
mainly responsible for the muons in this momentum range 
at the balloon altitude~\cite{hkkm-hamburg,gaisser-honda}.
Therefore, the study of the hadronic interaction model is for 
the primary cosmic rays in this energy region, corresponding to 
the neutrinos of 0.3--4~GeV.

For the wider energy region of primary cosmic rays, 
we may examine the hadronic interaction model using the observed muons
at different altitudes and at sites with different cutoff rigidities.
Note that at ground level the muon fluxes are available for a wider 
momentum range with good statistics.
We are preparing a paper~\cite{sanuki-mu} for such a study 
with the muon fluxes observed by BESS~\cite{motoki,sanuki-mountain,tanizaki},
and so limit ourselves here to comment that the muon flux observed by BESS is reproduced 
with DPMJET-III with an accuracy of $\sim$~5~\% for the muons
in the `important' momentum range from 1 to a few tens of GeV/c for 
most cases.
At ground level, the primary cosmic rays with energies 
from 20 to a few 100~GeV are responsible for the muons in this momentum
range, corresponding to neutrinos of 1--10~GeV.
This study was partly reported in Ref.~\cite{hkkm-hamburg}.

We do not use the original package of the hadronic interaction code
in the calculation of atmospheric neutrino fluxes.
We first carry out a computer experiment of the interaction of all kinds 
of primary or secondary cosmic rays with air-nuclei, using the
original hadronic interaction code.
Then, the `data' are used to construct an inclusive interaction code, 
which reproduces the multiplicities and 
energy spectra of secondary particles of the original code.
The inclusive interaction code violates the conservation laws for  
energy-momentum and other quantum numbers in a single interaction,
but they are restored statistically.
Note that 
for the secondary particles whose life time is shorter than $10^{-9}$~sec
we record their decay products as the data.
The experiment scans the energy region from 0.2 to $10^6$~GeV 
in kinetic energy, and is repeated typically 1,000,000 times for 
each kind of projectile and each injection energy.

For the energy distribution of secondary particles in the interaction,
we fit the original distribution of $x$, defined as $x \equiv E_k^{sec}/E_k^{proj}$, 
with the combination of B-spline functions for each kind of projectile particle,
each injection energy, and each kind of secondary.
Then the inclusive code uses the B-spline-fit to reproduce the energy distribution 
of the secondary particle with a good accuracy.
For the scattering angles, we calculate the average transverse momentum 
($<p_{\perp}>$) for each kind of projectile, 
each injection energy, each kind of secondary, 
and each secondary energy.
In the inclusive code, we sample the scattering angle ($\theta$) with
the distribution function $\propto \exp(a \cdot \cos\theta)\cdot d\cos\theta$, 
where $a$ is determined so that $<p_{\perp}>$ is the same as the 
original interaction model. 
The $p_{\perp}$-distribution approaches  
$\propto \exp(-a' \cdot p_{\perp}^2) \cdot p_\perp d p_\perp$ and
$a' = \pi/(2<p_{\vert}>)^2$ for $p \gg$~1~GeV/c.
Note, the inclusive code constructed for DPMJET-III reproduces 
not only $<p_{\perp}>$ 
but also, approximately, 
the original $p_\perp$-distribution for $p_\perp < 1$~GeV/c.
There is a longer tail in the original $p_\perp$ distribution 
for larger $p_\perp$.
However, since the number of secondary particles which 
have $p_\perp > 1$~GeV/c is limited, they are not important in this study.

The constructed inclusive codes are typically $\sim$100 times faster 
than the original package.
The fast computation is very important in the 3-dimensional 
calculation of the flux of atmospheric neutrinos, 
as well as the study of secondary cosmic rays.
Note, however, the inclusive interaction code is only valid for the 
calculation of a time averaged quantity, such as the fluxes of atmospheric 
neutrinos and muons.
The situation is similar to the superposition model for the nuclear 
cosmic rays.

\section{\label{sec:scheme}Calculation Scheme}

Except for the geomagnetic field model, 
the simulation scheme is similar to the previous 3-dimensional 
calculation~\cite{hkkm-dipole} 
in which we assumed a dipole geomagnetic field.
In this calculation, we use the IGRF geomagnetic field model~\cite{igrf} 
with the 10th order expansion of spherical functions for the year 2000.
As the geomagnetic field changes very slowly, the neutrino flux 
calculated for the year 2004 would not show a 
noticeable difference.
We use the US-standard 1976~\cite{us_standard}
atmospheric model, as in the previous study.
Note that
for a study of the seasonal variations of atmospheric neutrino fluxes
we need to use a more sophisticated and detailed atmospheric 
model~\cite{nrlmsise00}.

We assume the surface of the Earth is a sphere with radius of $R_e=6378.180$~km.
We also assume 3 more spheres; the injection, simulation, and escape spheres.
The radius of the injection sphere is taken as $R_{inj}= R_e + 100$~km,  
the simulation sphere as $R_{sim}= R_e + 3000$~km, 
and the escape sphere as  $R_{esc}= 10 \times$$R_e$.
The sizes of the injection sphere ($R_{inj}$) and escape sphere ($R_{esc}$)
are the same as in the previous study~\cite{hkkm-dipole}.

The cosmic rays are sampled on the injection sphere uniformly toward inward directions, 
following the given primary cosmic ray spectra. 
Before they are fed to the simulation code for propagation in air, 
they are tested to determine whether they can pass the rigidity cutoff, 
i.e., the geomagnetic barrier.
For a sampled cosmic ray, the `history' is examined by solving the 
equation of motion in the negative time direction.
When the cosmic ray reaches the escape sphere without touching the injection
sphere again in the inverse direction of time, 
the cosmic ray can pass through the magnetic barrier following the trajectory 
in the normal direction of time.
In the 1-dimensional calculation we normally prepare a cutoff table 
for each neutrino detector site beforehand, 
but it is practically impossible to construct such a table for 
the 3-dimensional calculation.
Note, all the nucleons carried by the cosmic ray nuclei are treated 
as protons with double rigidity ($=$ momentum for protons), 
before the first interaction with an air-nucleus and in 
the rigidity cutoff test.

The propagation of cosmic rays is simulated in the space between 
the surface of Earth and the simulation sphere.
When a particle enters the Earth, it loses
its energy very quickly, and generates neutrinos with energy less than
100~MeV only.
Therefore, we discard such particles as soon as they enter the Earth, as
most neutrino detectors which observe atmospheric neutrinos do
not have sensitivity below 100~MeV.

For secondary particles produced in the interaction of a cosmic ray 
and air-nucleus, there is the possibility that they go out and re-enter 
the atmosphere and create neutrinos with energy $\sim$1~GeV.
Therefore, too small a simulation sphere may miss such secondary particles.
On the other hand, it is very time consuming to follow all the 
particles out to distances far from the Earth.
In the previous study, 
we took the radius of simulation sphere to be $R_{sim}= $$R_e + 300$~km, 
and showed that this is sufficient to calculate the neutrino 
flux to within a accuracy of $\sim$1~\%
from an analysis of the neutrino production time after the first 
interaction~\cite{nu2002}.
In this paper, however, we adopt a radius for the simulation sphere of
$R_{sim}= $$R_e + 3000$~km for greater accuracy, since
we found the average computation time for a primary cosmic ray does not increase 
that rapidly up to a simulation sphere of this size. 
Regarding the size of simulation sphere, we study the
neutrino production time after the injection of the primary cosmic ray
in Sec.~\ref{sec:prodtime}.

We `observe' the neutrino at the surface of the Earth, 
and the size of the `virtual detector' is closely related to 
the accuracy of the calculated flux and computation time.
With too large a virtual detector, the average observation conditions,
such as the dependence on the geomagnetic field,
may differ from the real site.
However, with too small a virtual detector, 
it is difficult to collect a sufficient number of neutrinos within 
a reasonable computation time.
In the previous study, we assumed an axial symmetry with dipole geomagnetic 
field, and considered a belt around the Earth as the virtual detector. 
For the more realistic geomagnetic filed model IGRF,
we consider a localized virtual detector, the surface of the Earth  
inside a circle with the radius of $\sim$~1117~km (center angle of $10^\circ$) 
around the target detector. The virtual detector is $\sim 1/6$ the size of
the previous one. 

Note, we placed many virtual detectors on the Earth corresponding to the 
existing neutrino detectors, 
and recorded neutrinos for each detector at the same time.
However, we only show the results for the virtual detectors placed at 
Kamioka and North America in this paper,
as they are good examples of a low magnetic latitude
and a high magnetic latitude, respectively.
The fluxes for Soudan and Sudbury are almost identical,
and we refer to them here as North America.

\subsection{\label{sec:prodtime}Neutrino production time.}

Before showing the resulting atmospheric neutrino flux,
we would like to introduce some interesting quantities;
the neutrino production time and the impact parameter of primary cosmic rays.
These quantities provide important hints for the efficient calculation 
of atmospheric
neutrino fluxes in the 3-dimensional scheme.

First, we show the study of neutrino production time.
Before the calculation of atmospheric neutrino fluxes,
we studied the neutrino production time to optimize the size of 
simulation sphere.
Note, the radius of the simulation sphere used in this study is 
$R_{sim}$ = 10 $\times$ Re = 63,781.80~km,
and so neutrinos produced within 
$t_{free}=2\times (R_{sim} - R_{inj})/c \sim 0.4$~sec
after the injection of cosmic ray at the injection sphere
are absolutely free from the boundary, 
by the naive discussion of the causality.

\begin{figure}[tbh]
\includegraphics[width=10cm]{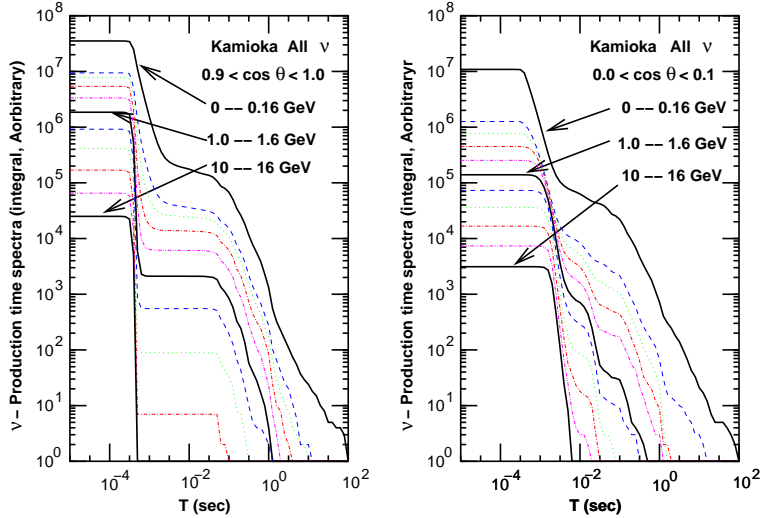}%
\caption{\label{fig:time-kam}
Integral production time distribution for the atmospheric neutrinos observed at Kamioka.
For the neutrino energies, we divide each decade into 5 bins, and 
show the production time spectra in solid, dashed, dash-dot, dash-dot-dot, and
dash--3-dot lines.
}
\end{figure}

We show the integral distribution of the neutrino production time 
for neutrinos observed in Kamioka in Fig.~\ref{fig:time-kam}.
The production time is measured after the injection of the cosmic ray 
at the injection sphere.
We find that  
$t_{free}$ for $R_{sim}= $$R_e + 3000$~km ($\sim$~20~msec) 
is large enough to calculate the flux of atmospheric neutrinos with an accuracy 
much better than 1\%.
It is interesting that there is a second peak at $\sim$0.1~sec due to the 
albedo particle reported by AMS.

We used $R_{sim}= $$R_e + 300$~km in the previous study~\cite{hkkm-dipole},
and reported that this $R_{sim}$ is sufficient for calculation of 
atmospheric neutrino fluxes with an  accuracy of $\sim$1\%~\cite{nu2002}.
Note there is a $\sim$0.3~msec time offset in the production time after the
injection for vertical directions,
and $\sim$3~msec time offset for horizontal directions.
Considering that the offset time is different for each injected 
cosmic ray, 
the study of the production time after the first interactions of cosmic rays
requires a more sophisticated study than the present one.
We can expect much better accuracy with $R_{sim}= $$R_e + 3000$~km.

The same study has been carried out for other neutrino detector sites.
However, the distributions are similar to that of Kamioka
except for the height and position of the second peak.
The second peak is lower and at larger production time for the 
higher latitude site.

\subsection{\label{sec:impact}Impact parameter of the primary cosmic rays.}

\begin{figure}[tbh]
\includegraphics[width=15cm]{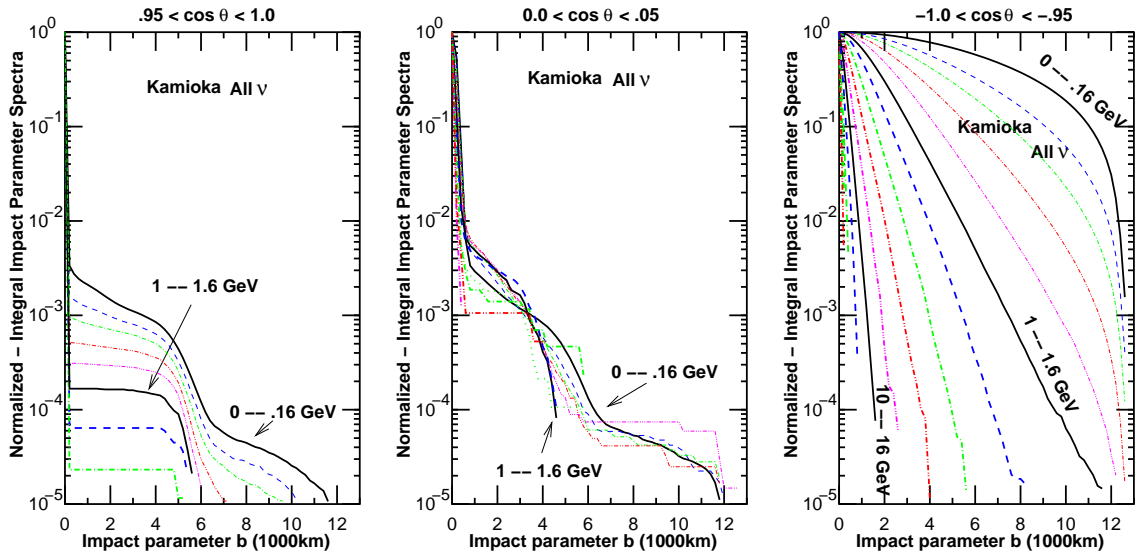}%
\caption{\label{fig:impact-kam}
Normalized integral impact parameter distribution of the primary cosmic rays
which produce neutrinos observed at Kamioka.
Each decade of neutrino energy is divided into 5 bins, and 
the impact parameter spectra are shown 
by solid, dashed, dash-dot, dash-dot-dot, and
dash--3-dot lines.
}
\end{figure}

Here, we study the impact parameter of cosmic rays
which produce a neutrino passing through the detector.
When the cosmic ray produces a neutrino going through a virtual detector, 
the impact parameter ($b$) is calculated against 
the contact point of the neutrino and the surface of 
Earth for the input primary cosmic ray at the injection sphere.

The impact parameter distributions are normalized and 
integral form are shown for Kamioka in Fig.~\ref{fig:impact-kam}.
As expected, the impact parameters 
of downward going neutrinos are distributed in a narrow region
near $b=0$,
while those of upward going neutrinos are widely distributed.
However, we find the distribution shrinks to $b=0$ as
the neutrino energy increases, and
for neutrinos with energy $>$~10~GeV, most neutrinos ($>$ 99\%)
are produced by primary cosmic rays with $b <$~1000~km.

The study of impact parameters can be used to accelerate 
the calculation of the atmospheric neutrino flux above $\gtrsim$~10~GeV.
Selecting an impact parameter for primary cosmic rays of $<$~2000km
at the injection results in
atmospheric neutrino flux calculations $\sim$30 times faster than 
the original calculation scheme explained above.
We use this acceleration technique for the neutrino flux above 10~GeV.

In Kamioka, it is found that the impact parameter distribution has
a structure at 5--6000~km for the vertical directions, and 3--5000~km
for horizontal directions.  This is considered again to be the effect
of the albedo particles observed by AMS.  However, the contribution is
small ($\ll$~1\%).  
The same study has been done for other neutrino detector sites.
However, the concentration of the impact parameter distribution to $b=0$
is quicker than Kamioka for downward going neutrinos, as they site 
at higher geomagnetic latitudes than Kamioka.
For upward going neutrinos, the impact parameter distribution is almost the
same for all the sites.

\section{\label{sec:flux}The flux of atmospheric neutrinos}

Without limiting the impact parameter,
we sampled 307,618,204,971 cosmic ray nucleons before the rigidity cutoff test, and 
simulated the propagation in air for 116,086,900,000 nucleons 
with kinetic energy $>$~1~GeV, 
or equivalently all the cosmic rays with $E_k/A > 1$~GeV 
arrive on the injection sphere in $8.07\times 10^{-8}$~second. 
Limiting the impact parameter,
we sampled 415,711,823,606 cosmic ray nucleons
before the rigidity cutoff and impact parameter test, 
and simulated the propagation in air for 25,413,045,195 nucleons 
with kinetic energy $>$~10~GeV 
or equivalently all the cosmic rays with  $E_k/A > 10$~GeV 
arrive on the injection sphere in 1.4 micro second.
Note the flux tables for Kamioka, Gran Sasso, and North America
calculated in this study are available 
at \url{http://www.icrr.u-tokyo.ac.jp/~mhonda}.

In this section we present the characteristic features of the 
atmospheric neutrino flux calculated in the 3-dimensional scheme (3D) 
and compare them with those calculated in the 1-dimensional scheme 
with the same primary cosmic ray flux and interaction model. 
To study the differences due to the interaction model and
the calculation scheme in 3-dimensional calculation, 
we also compare the atmospheric neutrino flux calculated in Ref.~\cite{fluka-battis}
(FLUKA),
and the one calculated in our previous 3-dimensional study
with the dipole magnetic field~\cite{hkkm-dipole} (DIPOLE).
Note, the interaction model used in DIPOLE is the same as Ref.~\cite{hkkm95}.
Interaction models and the geomagnetic field models used in the calculations 
are summarized in Table~\ref{tab:calsummary}.

For Kamioka and Gran Sasso, we calculated the atmospheric neutrino 
fluxes considering the effect of the surface structure (mountains) above 
the neutrino detectors.
However, in the following studies, we use the neutrino flux calculated 
for a flat detector at  sea level, i.e., ignoring surface structure,
to see the differences due to the calculation schemes.

\begin{table}[!htb]
\caption{Interaction and geomagnetic field models in the calculations.}
\vspace*{2truemm}
\begin{tabular}{|c| c c c |}
\toprule
Calculation &\hspace{3mm}Int. Model\hspace{3mm}&\hspace{3mm}Geomagnetic Field\hspace{3mm}&\hspace{3mm}Geomagnetic field\hspace{3mm} \\
 &  & \hspace{3mm}(Rigidity cutoff)\hspace{3mm} &  (In air) \\
\hline
3D & DPMJET-III & IGRF & IGRF \\
FLUKA\cite{fluka-battis} & FLUKA\footnote{Advanced version from FLUKA97} & IGRF & None \\
DIPOLE\cite{hkkm-dipole} &\hspace{3mm} Fritiof 1.6 base\hspace{3mm} & Dipole & Dipole \\
\hline
1D & DPMJET-III & IGRF & none \\
\botrule
\end{tabular}
\label{tab:calsummary} 
\end{table} 

\subsection{\label{sec:general}Zenith angle dependence of the neutrino flux}

\begin{figure}[tbh]
\includegraphics[width=15cm]{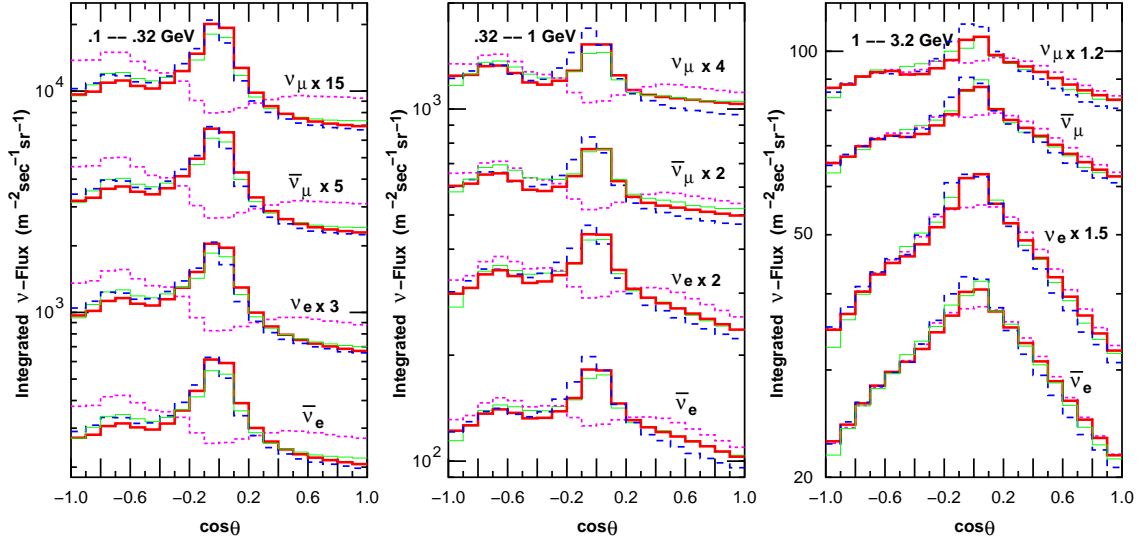}%
\caption{\label{fig:kam-zdep}
Zenith angle dependence of the atmospheric neutrino flux at Kamioka in 
3 energy bins. For the azimuthal directions, averages are taken.
The thick solid lines are for 3D,
dotted lines for 1D, dashed lines for FLUKA, and thin solid lines for 
DIPOLE.
}
\end{figure}

\begin{figure}[tbh]
\includegraphics[width=15cm]{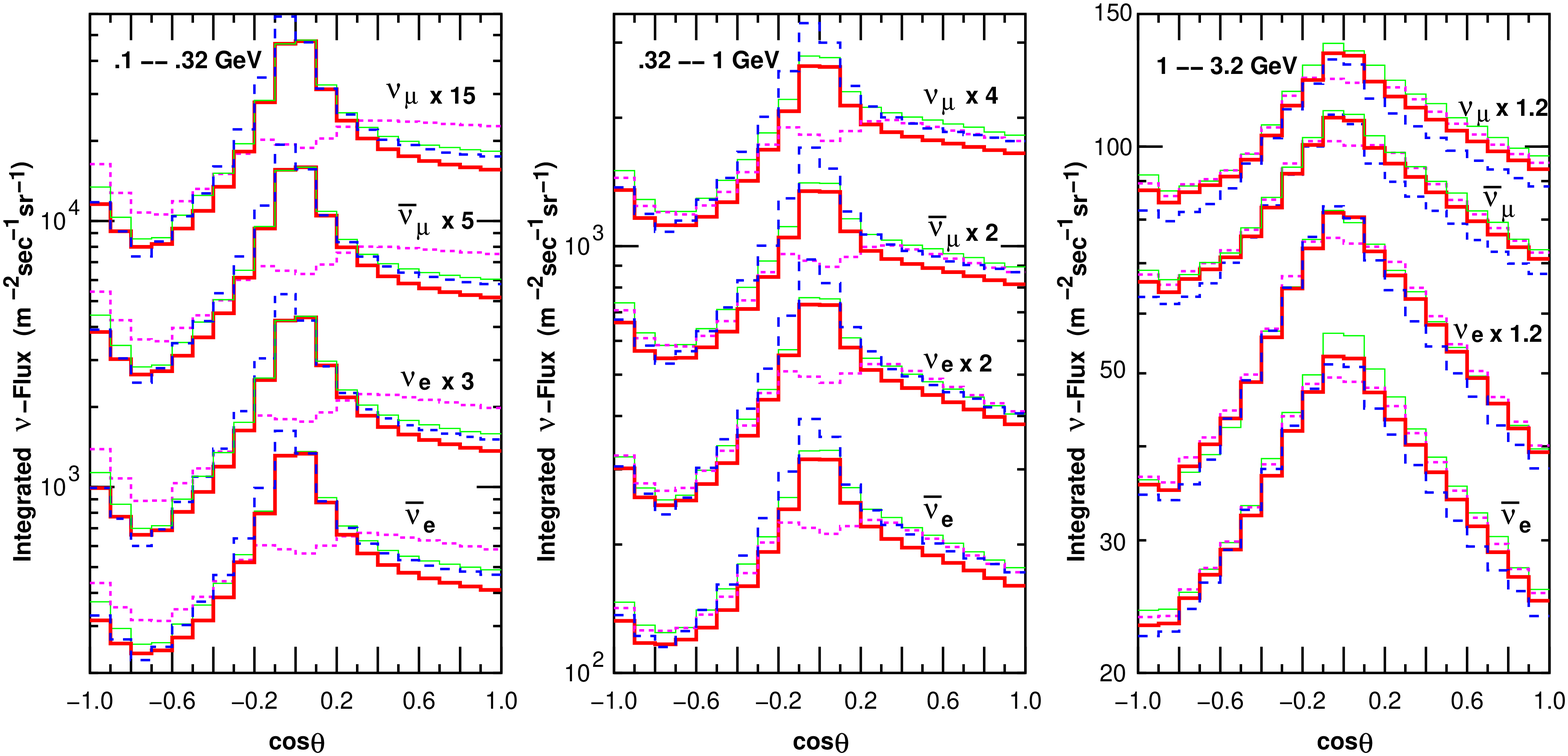}%
\caption{\label{fig:sno-zdep}
Zenith angle dependence of the atmospheric neutrino flux at North America in 
3 energy bins.
For the azimuthal directions, averages are taken.
The notation for lines is 
the same as Fig.~\ref{fig:kam-zdep}.
}
\end{figure}

The most prominent difference between 3-dimensional and 1-dimensional atmospheric
neutrino flux calculations is the horizontal enhancement at low energies.
(For the origin of the horizontal enhancement, see 
Refs.\cite{gaisser-honda,lipari-ge,hkkm-dipole})
We compare the zenith angle dependences of atmospheric neutrino fluxes 
calculated in the 3D, 1D, FLUKA, and DIPOLE cases
for Kamioka (Fig.~\ref{fig:kam-zdep}) and North America (Fig.~\ref{fig:sno-zdep}),
integrating over several energy bins and averaging over azimuth angles. 

\begin{figure}[tbh]
\includegraphics[width=15cm]{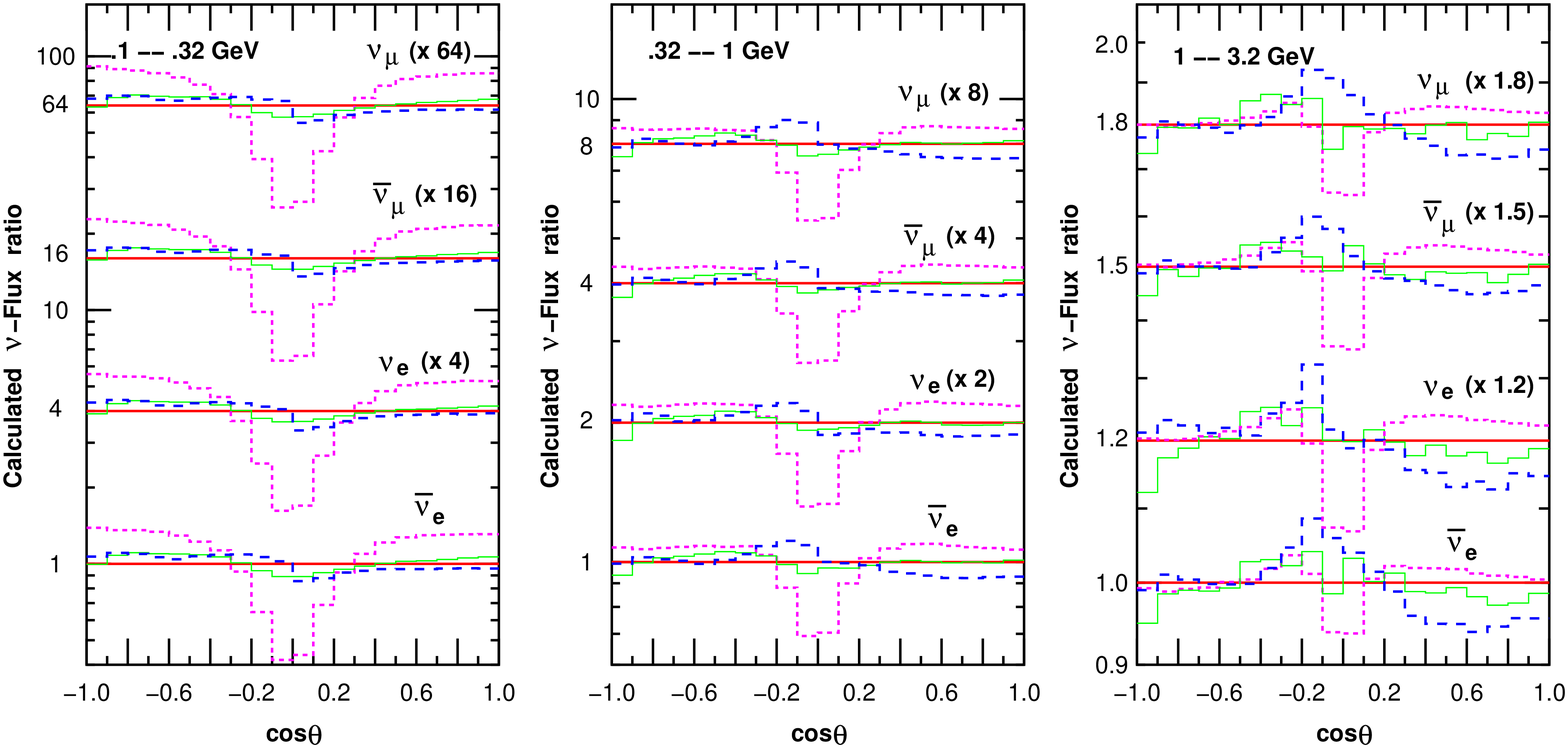}%
\caption{\label{fig:kam-zratio}
The normalized ratio of each flux to the 3D case as a function of zenith angle,
for Kamioka. 
Dotted lines are for 1D, dashed lines for FLUKA, and thin solid lines for 
DIPOLE.
}
\end{figure}

\begin{figure}[tbh]
\includegraphics[width=15cm]{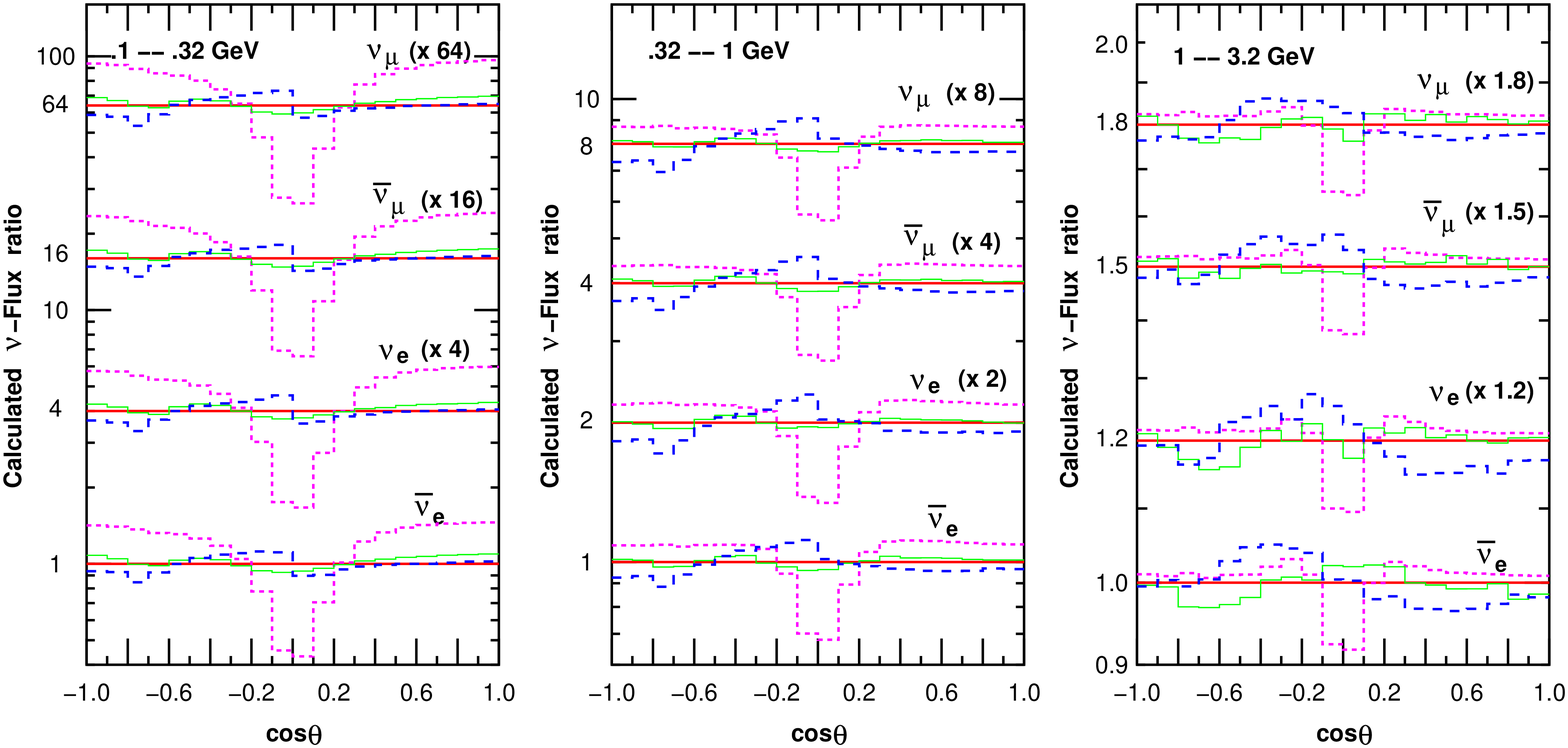}%
\caption{\label{fig:sno-zratio}
The ratio of atmospheric neutrino fluxes calculated by 1D, FLUKA, 
and DIPOLE for North America to that from the 3D case.
The notations are the same as for Fig.~\ref{fig:kam-zratio}.
}
\end{figure}

In these figures, we see the horizontal enhancements in 
0.1--0.32~GeV and 0.32--1~GeV energy bins for all the 3-dimensional calculations.
On the other hand, the flux near horizontal directions is rather smaller than 
neighboring directions in the 1D case due to the high cutoff rigidities.
In the 1--3.2~GeV energy bin, the differences between the calculations are small.
To study the difference of zenith angle dependences of neutrino fluxes 
due to the calculation scheme, 
we normalize each flux by the omni-directional flux average
and depict the ratio to the 3D case in Figs~\ref{fig:kam-zratio}
and \ref{fig:sno-zratio}
as a function of zenith angle.

In these figures, we find that the horizontal enhancement still 
exists in the 1--3.2~GeV energy bin, but that it decreases rapidly with 
neutrino energy.
The difference at near horizontal directions is more than 50~\% in 
0.1--0.32~GeV bin, but it reduces to $\lesssim$~10\% in 1--3.2~GeV bin,
for all kinds of neutrino.

The differences among the 3-dimensional calculations (3D, FLUKA, DIPOLE) are small,
especially that between 3D and DIPOLE.
However, the amplitude of the horizontal enhancement in DIPOLE is
clearly slightly smaller than that in the 3D calculation.
This is thought to be due to the difference of interaction model,
especially to the average transverse momentum of secondary mesons.
Note, the $<P_t>$ of pions is 0.289~GeV$/$c in DPMJET-III, 
while it is 0.256~GeV$/$c in Fritiof~1.6 for P~+~Air interactions 
at $E_{lab}=$10~GeV.
The difference between 3D and FLUKA is asymmetric below and above
the horizontal direction ($\cos\theta = 0$).
It is difficult to deduce differences in the interaction model 
in this comparison.

\begin{figure}[tbh]
\includegraphics[width=10cm]{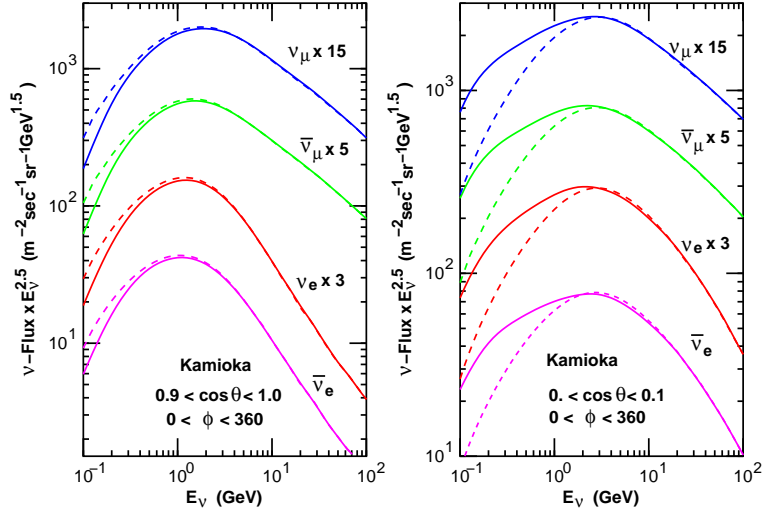}%
\caption{\label{fig:kam2dir}
Atmospheric neutrino fluxes at Kamioka for 3D and 1D, averaged over azimuth angles.
The left panel shows the comparison for vertical directions,
and the right panel for horizontal directions.
The solid lines show the 3D results and the dashed lines the 1D results. 
}
\end{figure}

\begin{figure}[tbh]
\includegraphics[width=10cm]{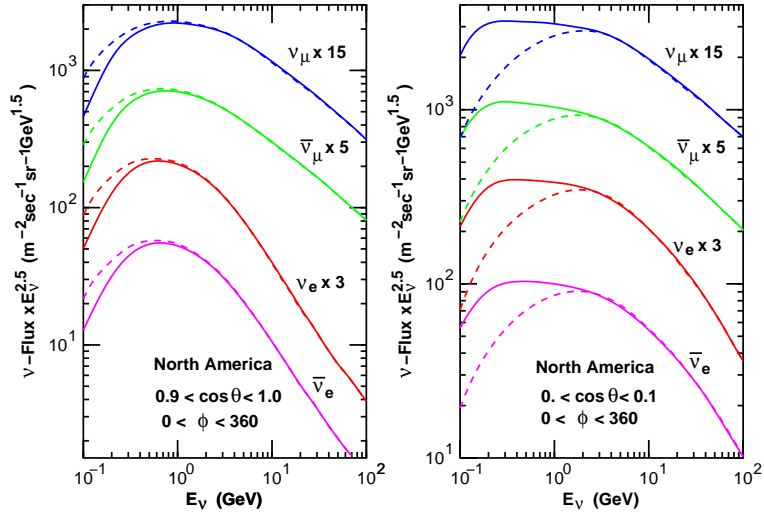}%
\caption{\label{fig:sno2dir}
Atmospheric neutrino fluxes at North America for 3D and 1D, averaged over azimuth angles.
The left panel shows the comparison for vertical directions,
and the right panel for horizontal directions.
The solid lines show the 3D results and the dashed lines the 1D results. 
}
\end{figure}

To see the energy dependence of the horizontal enhancement
more clearly, we compared the 3D and 1D energy spectra 
for vertical and horizontal directions averaging over azimuth angles 
in Figs.~\ref{fig:kam2dir} (Kamioka) and \ref{fig:sno2dir} (North America).
We find the differences disappear at $\sim$1~GeV for 
vertical directions, 
and $\sim$3~GeV for horizontal directions, for all neutrinos.

Moreover, we find the fluxes averaged over all directions for 3D and 1D cases
are very close to each other, even at low  energies.
Averaging the neutrino fluxes in 
Figs ~\ref{fig:kam-zdep} and \ref{fig:sno-zdep} over zenith angles,
the 3D/1D ratios are tabulated in table~\ref{tab:alldirratio}.
They agree with each other to within a few \% in all cases.

\begin{table}[!htb]
\caption{3D/1D ratio for the all-direction average of atmospheric neutrino flux.}
\vspace*{2truemm}
\begin{tabular}{|c| l l l l |}
\toprule
\hspace{7mm}E$_\nu$(GeV)\hspace{7mm} &\hspace{7mm}$\nu_\mu$\hspace{7mm}&\hspace{7mm}$\bar\nu_\mu$\hspace{7mm}&\hspace{7mm}$\nu_e$\hspace{7mm}&\hspace{7mm}$\bar\nu_e$\\ 
\hline
\multicolumn{5}{|c|}{Kamioka} \\
\hline
0.1 -- .32 & 0.979 & 0.980 & 0.970 & 0.978 \\
.32 -- 1.0 & 0.997 & 1.000 & 0.999 & 0.992 \\
1.0 -- 3.2 & 0.983 & 0.984 & 0.982 & 0.975 \\
\hline 
\multicolumn{5}{|c|}{North America} \\
\hline
 0.1 -- .32 & 1.036 & 1.035 & 1.028 & 1.025 \\
 .32 -- 1.0 & 1.019 & 1.020 & 1.021 & 1.014 \\
 1.0 -- 3.2 & 0.992 & 0.990 & 0.989 & 0.985 \\
\botrule
\end{tabular}
\label{tab:alldirratio} 
\end{table} 

\subsection{\label{sec:eastwest}East--West effect}

Here, we use the 3D and 1D fluxes only, since they are calculated 
under the same conditions 
(except for the 1- or 3-dimensional calculation scheme).
Contrary to the quantitative agreements between 3D and 1D above a few GeV
in the azimuthally averaged fluxes,
they are quite different when the azimuth angles are limited to 
East or West directions, even at higher energies ($\sim$~10~GeV).

\begin{figure}[tbh]
\includegraphics[width=10cm]{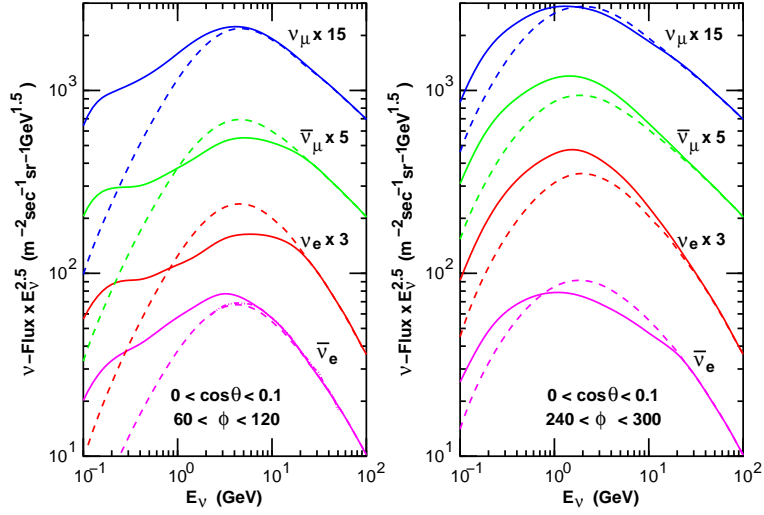}%
\caption{\label{fig:kam-horizontal}
The flux of horizontal ($0. < \cos\theta < 0.1$)
atmospheric neutrinos arriving from the East 
($60^\circ <\phi<120^\circ$, Left), 
and West ($240^\circ <\phi<300^\circ$, Right)
directions at Kamioka.
The solid lines show the result of the 3D calculation 
and the dashed line the 1D case. 
}
\end{figure}

\begin{figure}[tbh]
\includegraphics[width=10cm]{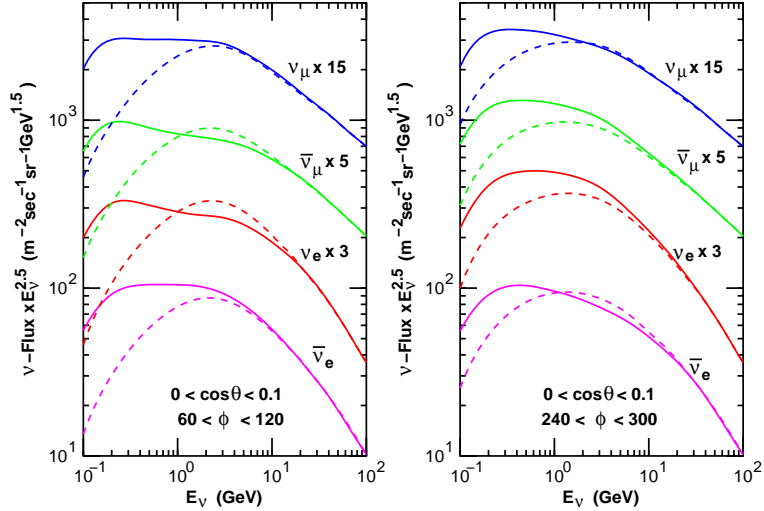}%
\caption{\label{fig:sno-horizontal}
The flux of horizontal ($0. < \cos\theta < 0.1$)
atmospheric neutrinos arriving from the East 
($60^\circ <\phi<120^\circ$, Left), 
and West ($240^\circ <\phi<300^\circ$, Right)
directions at North America.
The solid lines show the result of the 3D calculation 
and the dashed line the 1D case. 
}
\end{figure}

We depict the 3D and 1D  atmospheric neutrino fluxes
arriving horizontally ($0. < \cos\theta < 0.1$)
from the East ($60^\circ <\phi <120^\circ$),
and West ($240^\circ <\phi <300^\circ$) for 
Kamioka (Fig.~\ref{fig:kam-horizontal}) and North America
(Fig.~\ref{fig:sno-horizontal}), where
we measure the azimuth angle from South ($\phi=0$), and 
$\phi =90^\circ, 180^\circ, 270^\circ$ are East, North, and West
directions respectively.

The differences in the fluxes from the 3D and 1D calculations
for different kinds of neutrinos 
may be classified 
into two groups, $\nu_\mu$ and $\bar\nu_e$, and $\bar\nu_\mu$ and $\nu_e$. 
The former group are the decay products of
$\mu^-$, and the latter are the decay products of $\mu^+$.
The geomagnetic field deflects $\mu^+$'s toward the same direction as for
primary cosmic rays. 
Therefore, it enhances the East and West differences of the $\bar\nu_\mu$ and $\nu_e$ 
fluxes caused by the rigidity cutoff.
On the other hand, the geomagnetic field works on the $\mu^-$'s 
in the opposite direction to that for primary cosmic rays. 
Therefore, it reduces the East and West differences for the $\nu_\mu$ and $\bar\nu_e$ 
fluxes caused by the rigidity cutoff~\cite{lipari-ew}.
In the Figs.~\ref{fig:kam-horizontal} and \ref{fig:sno-horizontal}, 
we mainly see the horizontal enhancement in the neutrino energies 
$\lesssim$~1~GeV in the difference of 3D and 1D.
For $\gtrsim$ 1~GeV, however, the muon curvature in the geomagnetic field
is a larger effect than the horizontal enhancement, and this extends to 
several tens of GeV for near horizontal directions.

\begin{figure}[tbh]
\includegraphics[width=15cm]{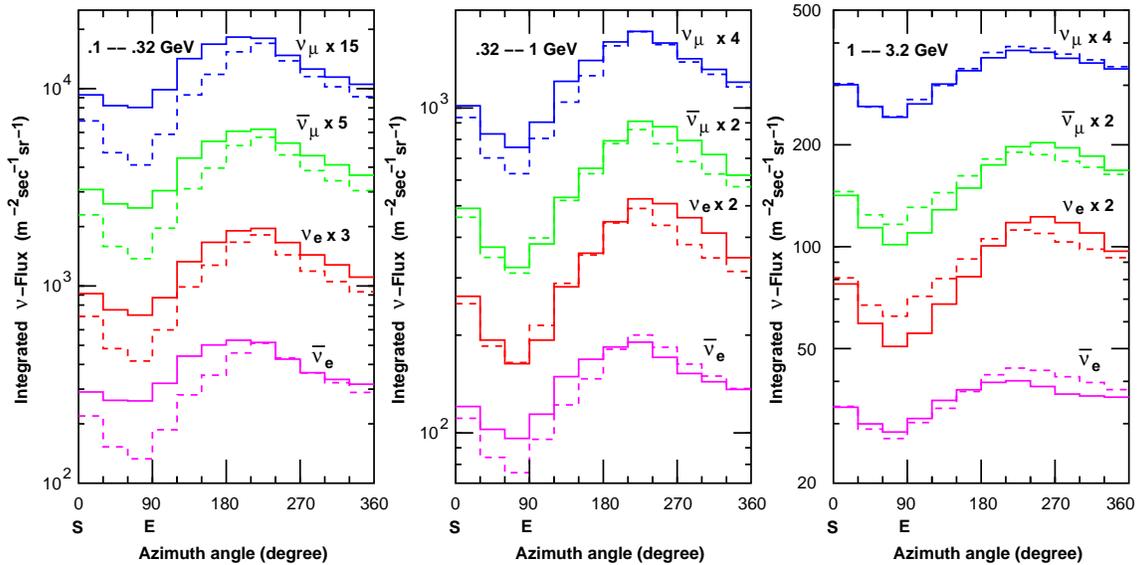}%
\caption{\label{fig:kam-adep}
Azimuth angle dependence of the atmospheric neutrino flux at Kamioka,
averaged over the zenith angle range $-0.5 < \cos\theta < 0.5$.
The solid lines show the 3D result and the dashed lines the 1D result. 
}
\end{figure}

\begin{figure}[tbh]
\includegraphics[width=15cm]{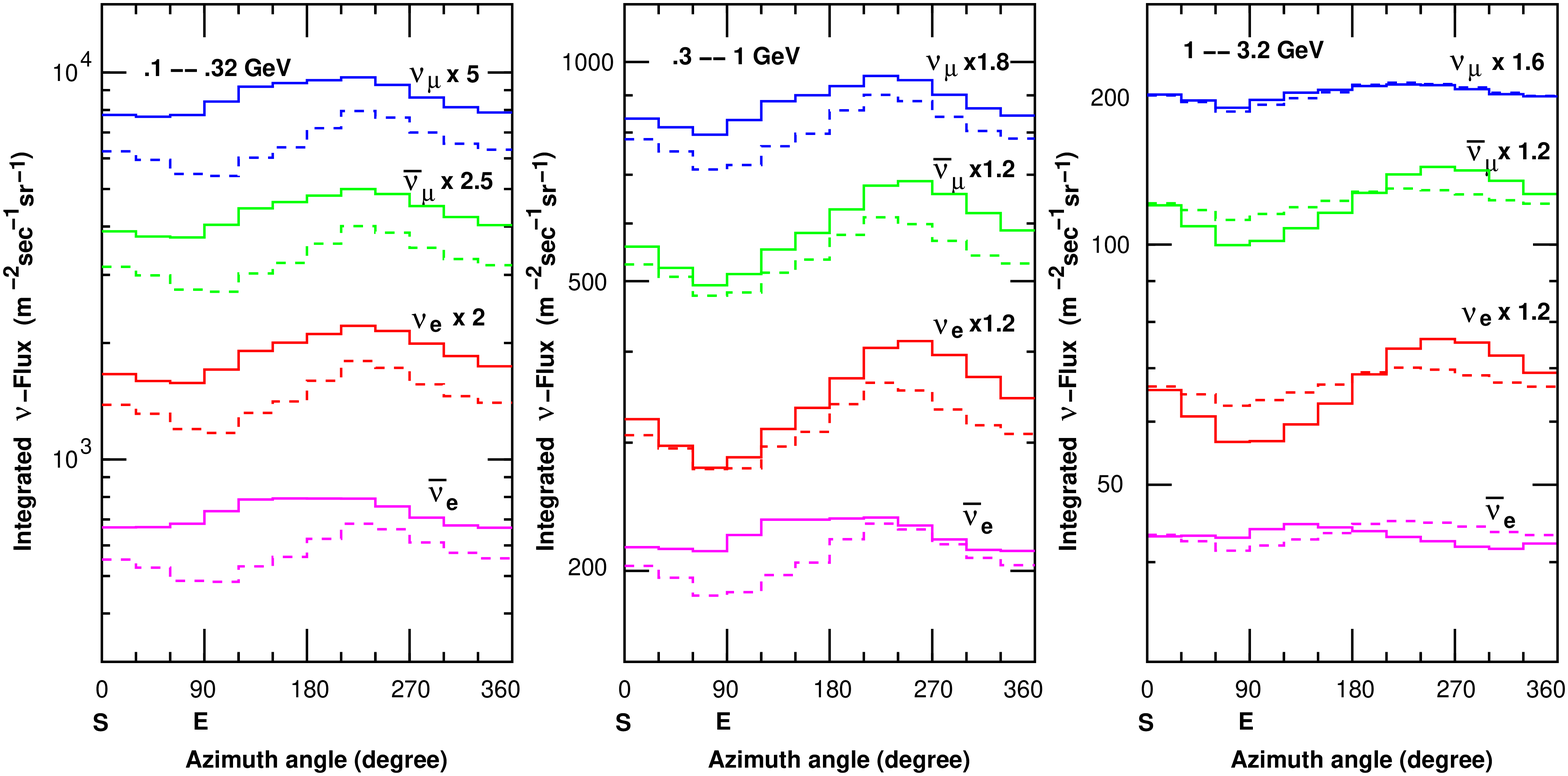}%
\caption{\label{fig:sno-adep}
Azimuth angle dependence of the atmospheric neutrino flux at North America,
averaged over the zenith angle range $-0.5 < \cos\theta < 0.5$.
The solid lines show the 3D result and the dashed lines the 1D result. 
}
\end{figure}

\begin{table}[!htb]
\caption{Max/Min ratio in Figs.~\ref{fig:kam-adep} and ~\ref{fig:sno-adep} 
as the amplitude of the azimuthal angle variation.}
\vspace*{2truemm}
\begin{tabular}{|c| l l l l | l l l l |}
\toprule
E$_\nu$(GeV) &$\nu_\mu$&$\bar\nu_\mu$&$\nu_e$&$\bar\nu_e$ &$\nu_\mu$&$\bar\nu_\mu$&$\nu_e$&$\bar\nu_e$
\\ 
\hline
&\multicolumn{4}{c|}{Kamioka, 3D}&\multicolumn{4}{c|}{Kamioka, 1D}
 \\
\hline
0.1 -- .32 & 2.27 & 2.51 & 2.75 & 2.03 & 4.13 & 4.13 & 4.36 & 3.85\\
.32 -- 1.0 & 2.27 & 2.82 & 3.22 & 1.98 & 2.73 & 2.76 & 2.98 & 2.65\\
1.0 -- 3.2 & 1.58 & 2.00 & 2.42 & 1.42 & 1.61 & 1.63 & 1.80 & 1.61\\
\hline 
&\multicolumn{4}{c|}{North America, 3D}&\multicolumn{4}{c|}{North America, 1D}
 \\
\hline
 0.1 -- .32 & 1.26 & 1.33 & 1.40 & 1.19 & 1.47 & 1.48 & 1.54 & 1.41\\
 .32 -- 1.0 & 1.20 & 1.39 & 1.49 & 1.11 & 1.27 & 1.28 & 1.31 & 1.26\\
 1.0 -- 3.2 & 1.07 & 1.25 & 1.35 & 1.07 & 1.08 & 1.09 & 1.12 & 1.09\\
\botrule
\end{tabular}
\label{tab:adep-amplitude} 
\end{table} 

The muon curvature effect should be seen in the azimuthal variation of the
atmospheric neutrino fluxes.
We show the integrated azimuthal angle dependence 
in the same energy bins as in Sec.~\ref{sec:general}
for Kamioka (Fig.~\ref{fig:kam-adep}) and North America (Fig.~\ref{fig:sno-adep}).
Also we tabulated the ratio of maximum to minimum fluxes in the figures
in Table~\ref{tab:adep-amplitude}, to see the variation amplitude.
In the 1D case, the amplitudes are similar to each other for all kinds of 
neutrinos in each energy bin. 
This is because the 1D azimuthal variation  is
caused by the rigidity cutoff of the primary cosmic rays.
In the 3D case, the amplitudes are different
among the different kinds of neutrino even in the same energy bin.
The amplitudes of $\nu_\mu$ and $\bar\nu_e$ are suppressed,
while those of $\bar\nu_\mu$ and $\nu_e$ are enhanced, except for the 
lowest energy bin of 0.1--0.32~GeV.
In the lowest energy bins, smearing suppresses the 
3D azimuth angle dependence.

Note that $\nu_e$ has the largest amplitude among all $\nu$'s
in 3D.
This is because about $1/2$ of the $\bar\nu_\mu$'s are created by 
pion decay directly, while all the $\nu_e$'s are created by $\mu$-decay
at these energies.
It is noteworthy that the amplitude of $\nu_e$ in the 1--3.2~GeV 
energy bin is still large.
This is important for the experimental confirmation of the effect of muon curvature, 
because the determination of the arrival direction is better for 
higher energy neutrinos.

Generally speaking, the azimuthal angle dependence of atmospheric neutrinos
at high magnetic latitude sites such as North America is smaller than that at
low magnetic latitude sites such as Kamioka because the rigidity cutoff is too low.
However, we find in Fig.~\ref{fig:sno-adep} and 
Table~\ref{tab:adep-amplitude} that
the difference in the azimuthal angle dependence of atmospheric 
neutrinos between 3D and 1D due to the muon curvature is similar to that for
the low magnetic latitude site.

\section{\label{sec:nflx-hi}Neutrino fluxes at higher energies}

\begin{figure}[tbh]
\includegraphics[width=10cm]{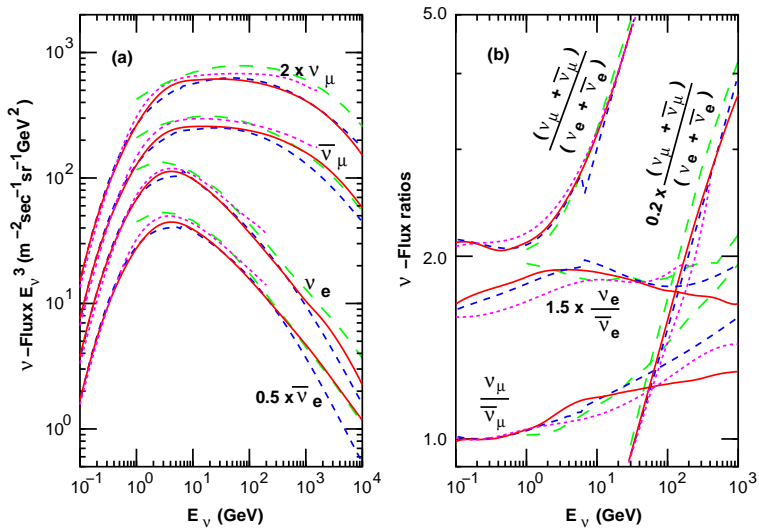}%
\caption{\label{fig:nflx-high}
(a) Atmospheric neutrino fluxes averaged over all directions. 
(b) Flux ratios $(\nu_\mu + \bar\nu_\mu)/(\nu_e + \bar\nu_e)$,
$\nu_\mu / \bar\nu_\mu$ and $\nu_e/\bar\nu_e$.
Solid lines are for this work, dotted lines for HKKM95,
dashed lines for FLUKA, and long dashed lines for BARTOL.
}
\end{figure}

In this section, we study the atmospheric neutrino flux at higher 
energies than in Sec.~\ref{sec:flux}.
First, we note that the atmospheric neutrino fluxes above 10~GeV have
a much larger uncertainty than those below 10~GeV.
The main reasons are the uncertainties in the primary cosmic rays and 
the hadronic interaction model at energies above 100~GeV.
As the difference between the neutrino fluxes 
calculated by 3-dimensional and 1-dimensional schemes are very 
small at the target energies here,
we include the 1-dimensional calculations from Refs.~\cite{hkkm95} (HKKM95) 
and \cite{gaisser-new} (BARTOL) in this comparison, and plot
them in Fig.~\ref{fig:nflx-high}.
Note that, as the 1D results are almost the same as the 3D case in this energy 
region, and are identical above 100~GeV, they are referred just as 
this work in this section.
Since the energy region available in the DIPOLE case is limited to below 10~GeV,
it is omitted from the comparisons.

The larger fluxes in HKKM95 and BARTOL are due to the larger
primary flux model used in HKKM95 (Fig.~\ref{fig:primary-proton}) 
and by the harder secondary spectrum in the hadronic
interaction model (TARGET-I) used in BARTOL.
We note that the ratios $(\nu_\mu + \bar\nu_\mu)/(\nu_e + \bar\nu_e)$
obtained for different calculations are very close.
The agreement among the different calculations is well within 5~\% 
at most energies.
However, the ratios $\nu_\mu / \bar\nu_\mu$ and $\nu_e/\bar\nu_e$
show larger differences.
The differences of $\nu_e/\bar\nu_e$ and  $\nu_\mu / \bar\nu_\mu$ above 
a few GeV are caused by differences in the pion and kaon productions 
and their charge ratios in interactions above a few  tens of GeV.
In particular,  $\nu_e/\bar\nu_e$ and  $\nu_\mu / \bar\nu_\mu$ above 100~GeV
are related to the kaon production in the hadronic interaction at energies
above 1~TeV.
Note it is still difficult to examine the hadronic interaction model
at these energies.

Next, we study the zenith angle variation of the atmospheric neutrino flux at 
energies $\gtrsim$~10~GeV with the quantity defined by

\begin{equation}
I_n (\cos\theta) = \int_{E1}^{E2} E_\nu^n {d N_\nu \over d E_\nu }(\cos\theta) dE_\nu\ .
\label{eq:stopnu}
\end{equation}
The neutrino interaction cross section increases 
approximately in proportion to the neutrino energy. 
Therefore, $I_1(\cos\theta)$ is approximately proportional to the rate of
neutrino events categorized as vertex contained and stop-muon events.
High energy muon neutrinos are also observed, arising from muons produced 
in the rock. 
In this case, the neutrino observation probability is proportional 
to the multiple of $[muon\ range] \times [neutrino\ cross\ section]$.
The muon range is approximately proportional to the muon energy 
below several 100~GeV, where the energy loss is dominated by ionization.
The event rate is approximately proportional to $I_2(\cos\theta)$.
Above $\sim$~500~GeV, the muon energy loss is dominated by 
radiative processes~\cite{paircreation}, and the relation fails.

\begin{figure}[tbh]
\includegraphics[width=10cm]{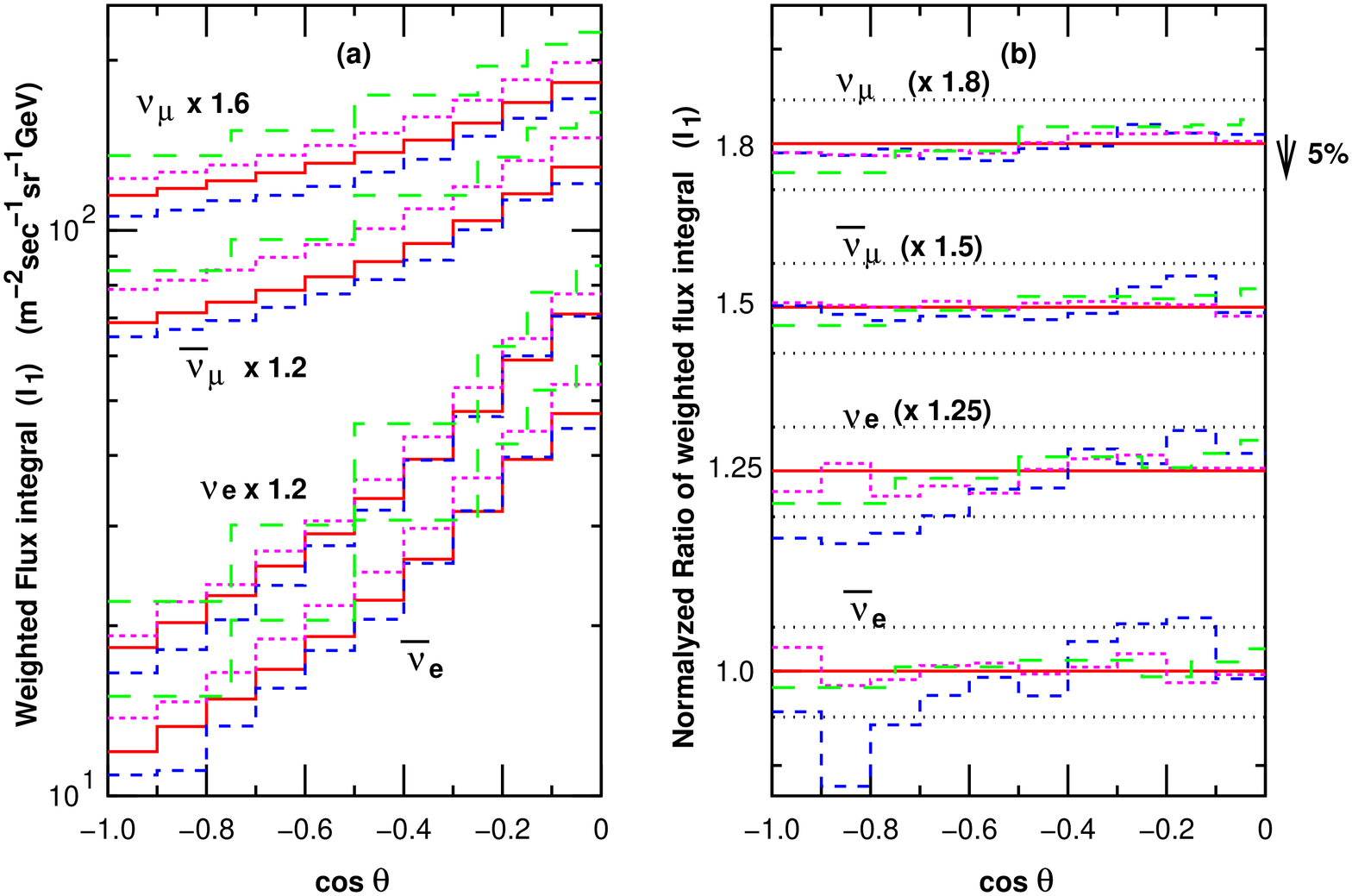}%
\caption{\label{fig:stopnu}
(a) Zenith angle variation of $I_1$ defined by Eq.\ref{eq:stopnu}.
(b) The normalized ratio of $I_1$ of each flux to this work as
a function of zenith angle. 
The solid lines are for this work, 
dotted lines for HKKM95, dashed lines for FLUKA,
and long dashed lines for BARTOL in both panels.
}
\end{figure}

$I_1(\cos\theta)$ is calculated with $E_1 = 3.2$~GeV and $E_2 = 1000$~GeV, 
and is shown in the left panel of Fig.~\ref{fig:stopnu} 
for this work, HKKM95, BARTOL and FLUKA.
Note, the median energy is $\sim$~6~GeV in this integration over 
all the fluxes.
We also depicted the normalized ratio of each flux to
this work in the right panel of Fig.~\ref{fig:stopnu}.
As the atmospheric neutrino flux is expected to be symmetric above
and below $\cos\theta=0$, we depicted the lower half ($\cos\theta < 0$) only.
Although there are large differences among theses calculations 
in the absolute values,
the differences of normalized fluxes are small, particularly for
$\nu_\mu$ and $\bar\nu_\mu$ ($\lesssim$~3~\%).

\begin{figure}[tbh]
\includegraphics[width=10cm]{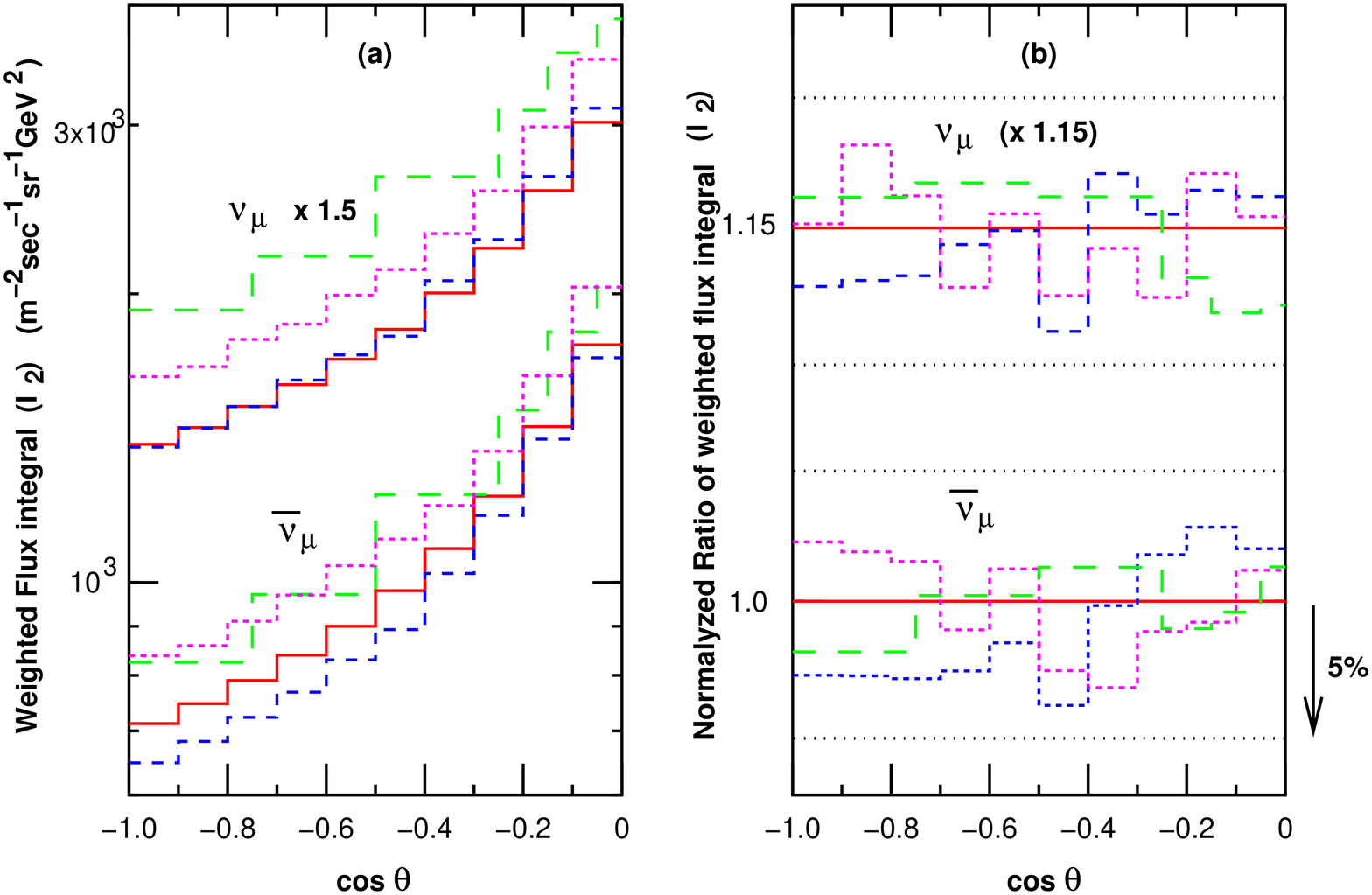}%
\caption{\label{fig:upnu}
(a) Zenith angle variation of $I_2$ defined by Eq.\ref{fig:stopnu}.
(b) Normalized ratio of $I_2$ of each flux to this work as
a function of zenith angle. 
The solid line is for this work, dotted line for HKKM95, dashed line for FLUKA,
and long dashed line for BARTOL in both panels.
}
\end{figure}

In Fig.~\ref{fig:upnu}, we show the zenith angle variation of $I_2$
with $E_1 = 10$~GeV and $E_2 = 1000$~GeV for this work, 
HKKM95, BARTOL, and FLUKA for $\nu_\mu$ and $\bar\nu_\mu$ fluxes. 
The median energy is $\sim$~100~GeV.
We find a large difference in absolute values as is expected 
from the left panel of Fig.~\ref{fig:nflx-high}.
However, the differences are small when they are normalized.
The ratio of the normalized weighted integral $I_2$ is shown
as a function of zenith angle in the right panel of Fig.~\ref{fig:nflx-high}.
The differences in normalized fluxes are $\lesssim$~3~\%.

As is seen in the left panel of Fig.~\ref{fig:nflx-high}, 
the ratios $\nu_\mu / \bar\nu_\mu$ and $\nu_e/\bar\nu_e$
differ significantly between the calculations above 10~GeV.
This is due to differences in the pion and kaon productions in the
interaction model used by the different calculations.
For example the multiplicity for kaons in DPMJET-III is almost 20\%
larger than that of FLUKA~97 at 1~TeV.
Note, the interaction model used in FLUKA is developed by the authors 
of FLUKA~97.
Despite these differences in the interaction models, 
the zenith angle dependences of atmospheric neutrinos show good 
agreement between the different calculations.

\section{\label{sec:height}Production height of atmospheric neutrinos.}

In the analysis of neutrino oscillations, the
distance from the point at which the neutrino originates to the detector 
plays an essential role in determining $\Delta m^2_{23}$.
To estimate the distance, the arrival zenith angle is used.
However, the arrival zenith angle does not determine the distance uniquely,
but gives a probability distribution for the distance.
In this section, we study the neutrino production height distribution 
for a given zenith angle.
As we assumed the surface of the Earth is a sphere,
the distance and production height are related by the formula:

\begin{equation}
d = \sqrt{(h^2+2R_e h) + (R_e \cos\theta)^2} - R_e \cos\theta~~,
\label{eq:h2d}
\end{equation}
where $h$ is the height, $R_e$ is the radius of the
Earth, and $d$ is the distance.
Note, the production height distribution changes its shape slowly as 
the zenith angle varies, while 
the distance distribution changes its shape
very quickly near horizontal directions.
Therefore, the study of production height distribution is more 
robust than the direct study of the distance distribution. 

In the experimental analysis,
the neutrino events are generally grouped in zenith angle bins 
irrespective of their azimuthal directions. 
Also, it is difficult to distinguish $\nu$ from $\bar\nu$ in 
current experiments.
Therefore, we study the production height averaged over the 
azimuthal angles and for the sum of  $\nu_\mu$ and $\bar\nu_\mu$, 
and $\nu_e$ and $\bar\nu_e$.

\begin{figure}[tbh]
\includegraphics[width=10cm]{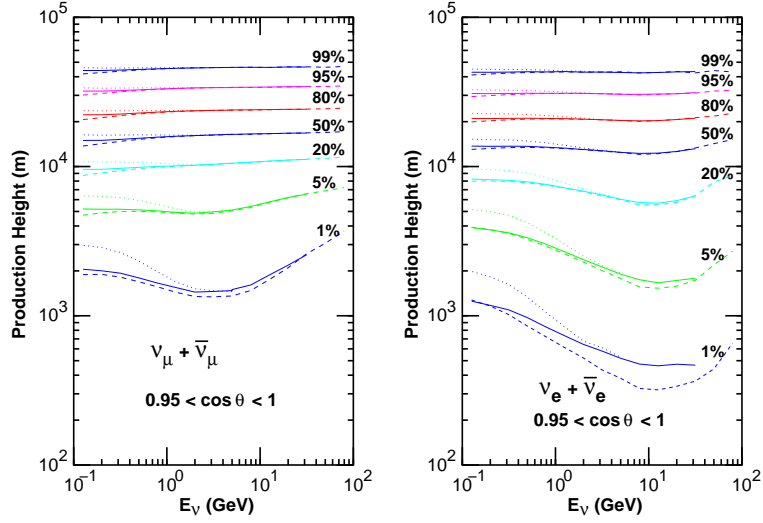}%
\caption{\label{fig:height-vertical}
The neutrino production height at near-vertical directions
for $\nu_\mu$ and $\bar\nu_\mu$ (left) and  $\nu_e$ and $\bar\nu_e$ (right).
The fixed accumulated probabilities of 
1~\% 5~\%, 20~\%,50~\%,80~\%, 95~\%, and 99~\% for the 
production height are shown as a function of the neutrino energy.
The solid lines are for 3D at Kamioka,
dashed lines for 1D at Kamioka,
and dotted lines for 3D at North America.
}
\end{figure}

\begin{figure}[tbh]
\includegraphics[width=10cm]{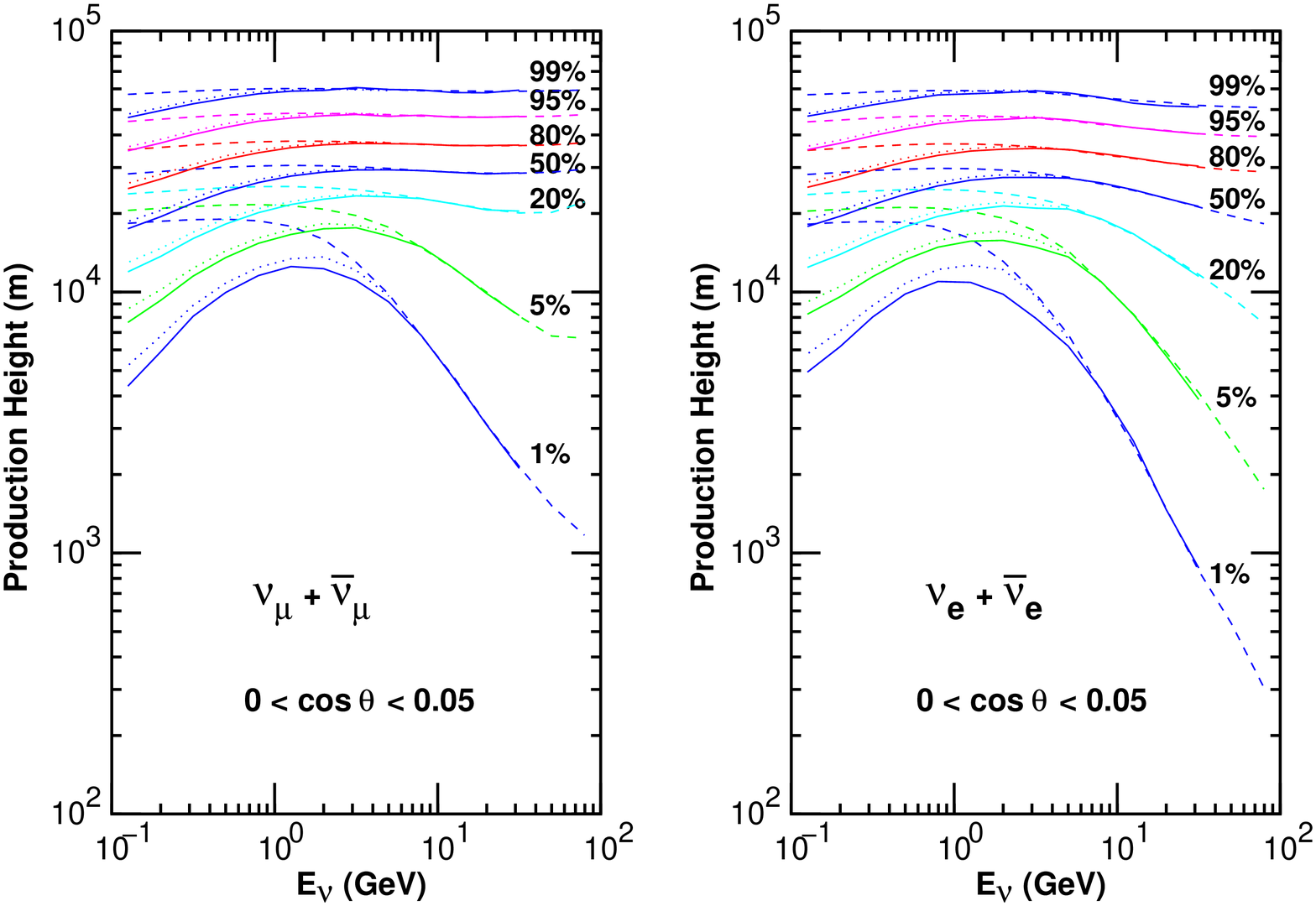}%
\caption{\label{fig:height-horizontal}
The neutrino production height at near-horizontal directions 
for $\nu_\mu + \bar\nu_\mu$ (left) and  $\nu_e + \bar\nu_e$ (right).
The fixed accumulated probabilities of 
1~\% 5~\%, 20~\%,50~\%,80~\%, 95~\%, and 99~\% for the 
production height are shown as a function of the neutrino energy.
The solid lines are for 3D at Kamioka,
dashed lines for 1D at Kamioka, and
dotted lines for 3D at North America.
}
\end{figure}

We show the neutrino production height distribution for vertical 
directions (Fig.~\ref{fig:height-vertical}) and for horizontal 
directions (Fig.~\ref{fig:height-horizontal}).
In the figures, we depict the lines for accumulated
probabilities of 1~\% 5~\%, 20~\%,50~\%,80~\%, 95~\%, and 99~\% for 
various neutrino energies.
Note we study the neutrino production height distribution in 
steps of $\Delta \cos\theta = 0.05$  for a better zenith
angle resolution. 

For the vertical directions, 
the production height in the 3D case is higher than that of 1D.
However, the differences are small and disappear at $\sim$~1~GeV 
for the median of the production height distribution.
Note the difference at the lower tail (1~\% line) 
exists even at a few tens of GeV due to the muon curvature.
On the other hand, for the horizontal directions the 3D production 
height is lower than that of 1D.
The differences are larger than those for vertical directions,
and are $\sim$~10~\% for the median of the height distribution even at 1~GeV.
However, they agree with each other at several GeV and above.

The differences of the production height resulting from differences 
in magnetic latitude are small.
The neutrino production heights for low energy primary cosmic rays 
are generally higher than those of higher energy primary cosmic rays.
Therefore, the neutrino production height at high magnetic latitude
is a little higher than that at low magnetic latitude.

Note,
the neutrino production height distribution has an azimuthal variation, 
mainly due to the muon curvature in the geomagnetic field.
The production heights of $\nu_\mu$ and $\bar\nu_e$ coming from 
eastern directions are lower than those coming from western directions, 
and the production heights of $\bar\nu_\mu$ and $\nu_e$ coming from
eastern directions are higher than those coming from western directions.
The azimuthal variation of the median production height for the 
horizontal directions is $\sim\pm$10~\% for $\nu_e$ and $\bar\nu_e$,
and $\sim\pm$5~\% for $\nu_\mu$ and $\bar\nu_\mu$ at a few GeV, 
and they slowly decrease with the neutrino energy.

The production heights for $\nu_\mu$ and $\bar\nu_\mu$ are
a little different even for the vertical direction. 
Considering the fact that the major component of cosmic rays
is protons, a relatively larger number of $\nu_\mu$ are 
created by pions rather than muons, and a relatively larger 
number of $\bar\nu_\mu$ are created by muons than pions.
The decay products of muons are generally produced at lower altitudes 
than the decay products of pions.
This is also the reason why we see a larger muon curvature effect in 
$\bar\nu_\mu$'s than in $\nu_\mu$'s.
This results in a small difference in the production height for 
$\nu_\mu$ and $\bar\nu_\mu$. 
For $\nu_e$ and $\bar\nu_e$, this mechanism is not relevant,
and the production heights are almost the same.

\section{\label{sec:summary}Summary and discussion}

We have revised the calculation of the atmospheric neutrino flux,
according to recent developments in primary cosmic ray observations 
and hadronic interaction models.
We have also updated the calculations from a 1-dimensional scheme 
to a 3-dimensional one.
For the interaction model, we compared the available interaction models
with the secondary cosmic rays observed at balloon altitudes, 
and selected DPMJET-III as the preferred model for this study.
We have constructed an inclusive hadronic interaction code based on 
DPMJET-III for speed. 
The computation speed is very important in the 3-dimensional calculation.
We have processed 307,618,204,971 cosmic rays with $E_k/A > 1$~GeV
for lower energy neutrino fluxes,
and 415,711,823,606 cosmic rays with $E_k/A > 10$~GeV
for neutrino fluxes above 10~GeV.
Combining both simulations, the statistical error due to the Monte Carlo method
is negligibly small for neutrino energies below a few tens of GeV.

With the primary fluxes accurately determined by BESS and AMS below 100~GeV,
the DPMJET-III interaction model, and the fast 3-dimensional simulation code
for the cosmic ray propagation, we consider we have reduced the
uncertainty to $\sim$~10~\%  in the calculation of the atmospheric 
neutrino flux at energies below 10~GeV, since we could reproduce the 
muon fluxes at various altitude with good accuracy from 1 to a few 
10~GeV~\cite{abe2,sanuki-mu} in this calculation scheme.
However, for the atmospheric neutrino flux above 10~GeV, 
the uncertainties in the atmospheric neutrino fluxes are still large 
due to the uncertainties of the primary cosmic ray flux and interaction model
above 100~GeV.

The differences we find between the 1- and 
3-dimensional calculation schemes are similar 
to those we found in the previous study with a dipole geomagnetic field.
When we average the atmospheric neutrino flux over azimuthal 
angles, the fluxes calculated in the 3-dimensional scheme quickly converge
with those calculated in the 1-dimensional scheme at a few GeV.
With the larger statistics, however, it becomes clear that 
muon curvature affects the horizontal neutrino flux
even at energies $\gtrsim$~10~GeV,
while most other `3-dimensional effects' disappear at a few GeV.

In comparison with other calculations of atmospheric neutrino flux,
we find the zenith angle dependences of different calculations are very similar, 
although there are differences in the absolute values.
The remaining differences of the zenith angle dependence
at higher energies ($\gtrsim$ 10~GeV) are consistent with the uncertainty 
of kaon production in the hadronic interaction~\cite{noon2003}.
Therefore, we may conclude that the main reason for the remaining difference of
the zenith angle dependence is in the kaon production of 
hadronic interaction model used by different calculations.
Note, there are large differences in the interaction model used by the
different calculations, as is known from the large $\nu_e/\bar\nu_e$ 
and $\nu_\mu/\bar\nu_mu$ differences at higher energies.

The production height distributions in the 1- and 3-dimensional
calculation schemes are different depending on the arrival direction.
When they are averaged over azimuthal directions, they agree with each 
other well above a few GeV, except for a small distortion at the lower tail 
for the vertical directions.
This situation is similar to that of the flux value.
There are azimuthal variations of the production height due to the
muon curvature, however
it is difficult to study them in detail with the statistics of
this work.
However, for the experimental study of atmospheric neutrinos 
for neutrino oscillations,
the azimuthal variations are not important.

It is interesting that
the effect of albedo particles observed by AMS at satellite altitudes
is seen in the neutrino production time distribution. 
The contribution of such particles to the atmospheric neutrino flux is
a little higher for the low magnetic latitude site (Kamioka) than the
high magnetic latitude site (North America),
but small ($ \ll 1$~\%) for both sites.

We expect that the validity of the calculation scheme and the effect of 
the muon curvature will be confirmed by the observation of
the azimuthal variation of the neutrino events.
Although the statistics are still insufficient, 
the SuperKamiokande experiment observed a larger amplitude of the azimuthal 
variation for the e-like events 
than that for the $\mu$-like events~\cite{futagami,moriyama} as is predicted
in Sec.~\ref{sec:eastwest}. 
We would like to note that the muon curvature effect has been confirmed 
by the azimuthal variation of the muon flux with an amplitude almost 
the same as the value predicted in our calculation scheme~\cite{hanoi-mu}.

\section{Acknowledgments}
We are grateful to 
J.~Nishimura, A.~Okada, P.~Lipari, T.~Sanuki, K.~Abe, S.~Haino, Y.~Shikaze and S.~Orito
for useful discussions and comments. 
We thank P.G.~Edwards for a careful reading of the manuscript.
We also thank ICRR, the University of Tokyo, for the support.
This study was supported by Grants-in-Aid, KAKENHI(12047206), from
the Ministry of Education, Culture, Sport, Science and Technology 
(MEXT).

\bibliography{nflx2004}

\end{document}